\documentclass[onecolumn]{revtex4}
\usepackage{graphicx}
\usepackage{dcolumn}
\usepackage{color}
\usepackage{bm}
\usepackage{amsmath}
\usepackage{amsfonts}
\usepackage{amssymb}
\input epsf
\begin{document}

\title{\Large Higher curvature gravity at the LHC}

\author{\bf Sumanta Chakraborty \footnote{sumantac.physics@gmail.com}}

\affiliation{IUCAA, Post Bag 4, Ganeshkhind,
Pune University Campus, Pune 411 007, India}

\author{\bf Soumitra SenGupta \footnote{tpssg@iacs.res.in}}

\affiliation{Department of Theoretical Physics,
Indian Association for the Cultivation of Science,
Kolkata-700032, India}

\date{\today}

\begin{abstract}
We investigate brane-world models in different viable $F(R)$ gravity theories where the Lagrangian
is an arbitrary function of the curvature scalar. Deriving the warped metric for this model,
resembling Randal-Sundrum (RS) like solutions, we determine the graviton KK modes. 
The recent observations at the LHC, which constrain the RS graviton KK modes to a 
mass range greater than 3 TeV, are incompatible to RS model predictions. 
It is shown that the models with $F(R)$ gravity in the bulk 
address the issue which in turn constrains the $F(R)$ model itself.
\end{abstract}

\maketitle


With the observable universe being $(3+1)$ dimensional, there is room for
additional unobserved dimensions. Such extra-dimensional models are primary 
candidates to address the well-known problems of the unnatural 
fine-tuning of the Higgs mass against radiative corrections due to the large 
hierarchy between the electroweak and Planck scale.

In order to address the above issues, brane-world models were proposed 
\cite{Randall1,Randall2,Witten} where our Universe is a hypersurface 
embedded in a higher-dimensional bulk.
The RS model provides a geometric solution to the hierarchy problem via an
exponential warp factor whose magnitude is controlled by separation of the two
3-branes and the bulk cosmological constant. The RS model constitutes a self-consistent
description of our Universe with four-dimensional standard Friedmann cosmology
and Newtonian limit reproduced \cite{Langlois,Csaki}.

The RS model line element is given by,
\begin{equation}\label{braner1}
ds_{RS}^{2}=e^{-2kr_{c}\phi}\eta _{\mu \nu}dx^{\mu}dx^{\nu}+r_{c}^{2}d\phi ^{2}
\end{equation}
with Greek indices $\mu ,\nu ,\ldots$ run over $0,1,2,3$ and they refer to usual observed dimensions,
while the co-ordinate $\phi$ signifies the extra dimension which is spacelike of length $r_{c}$.
The constant $k$ is connected to bulk cosmological constant $\Lambda$ such that,
$k=\sqrt{-\Lambda /12M^{3}}$. This is of the order of Planck mass. The quantity $e^{-2kr_{c}\pi}$
is known as the warp factor. The slices $\phi =0$ and $\phi =\pi$ represents hidden
and visible branes (i.e., our observable Universe) respectively. Due to presence of
the warp factor a mass scale of the order of Planck scale, $M_{Pl}$ gets warped on the visible brane as,
\begin{equation}\label{braner2}
 M=M_{Pl}e^{-2kr_{c}\pi}
\end{equation}
For $kr_{c}\simeq 12$, we get $M \simeq 1 TeV$. Hence in this picture
the stability of the Higgs mass against large radiative correction is controlled
by the warped geometry of the five dimensional spacetime.

The 2-brane system can be stabilized by introducing a bulk scalar field \cite{Goldberger1,Chakraborty}.
To address the cosmological constant problem, various other models have 
been proposed which include the domain wall
scenario \cite{Rubakov1,Verlinde1} and self-tuning in large extra dimension
\cite{Krause,Nilles}. However, this brane-world model has a drawback; it
describes the visible brane, i.e., our Universe with negative tension, which is
intrinsically unstable. Also the choice of the flat metric ansatz on the 
3-brane is not consistent with the present day observed
small value of cosmological constant. The model is extended to include 
maximally symmetric spaces,
and a more general setup was proposed \cite{Sengupta}. In this
work we consider the RS solutions and their generalizations for $F(R)$ gravity in 
the bulk (for recent review on $F(R)$ gravity see \cite{Nojiri1}, see also
\cite{Nojiri2}, \cite{Nojiri3}). Higher derivative gravity and the fine-tuning problem
have been discussed in Ref. \cite{Parry}, while Ref. \cite{Balcerzak} studies the cosmological
aspects of the $F(R)$ brane world. Some other features, such as effective Einstein equations
and junction conditions for $F(R)$ brane-world, were studied in \cite{Silva1} and \cite{Silva2}.

Our work has been inspired by experimental data from the LHC in the 
context of the search for extra warped dimensional models like RS. 
The search for a warp geometry model at the LHC is performed 
through the signal of the dilepton or diphoton 
decay channel of the first graviton KK modes and indicates the lower 
bound on the masses of the lowest graviton KK modes, greater than 
3 TeV (see for example \cite{Sengupta2} and \cite{LHC}). 
This poses serious problem to the standard RS model, since it is difficult to explain this large
lower bound within its theoretical framework. To elaborate this point we mention that 
3 TeV graviton can be accommodated in RS model by increasing value for $k$, which 
takes k above the Planck scale making the five-dimensional 
classical solution of Einstein's equation invalid. 
The other option will be to decrease $r_{c}$, keeping k fixed. This in turn increases 
the Higgs mass much above 126 GeV.
In this work we try to introduce a
possible generalization of the RS model to address this issue 
within the framework of the warped geometry model.

Recently there has been a boost in the gravitation research along the line of
modifying the gravitational Lagrangian in order to explain the high-energy
nonrenormalizable behavior of gravitation. These modified theories include
$F(R)$ theories, Gauss-Bonnet gravity, Lanczos-Lovelock models \cite{Deser} etc.
In this paper we shall study the warped geometric model with $F(R)$
Lagrangian in the bulk and try to find out the graviton KK modes in 
these warped geometric models. Though the $F(R)$ model contains 
higher derivative terms of the metric, for $R=\textrm{constant}$ hypersurface all 
these higher derivative terms vanish \cite{Nojiri1}. 
Here we exclusively work on this constant curvature slice such that 
higher derivative terms do not appear.

We start from the following action for $F(R)$ gravity on the bulk,
\begin{equation}\label{braneg1a}
S=\int d^{5}x \sqrt{-G}\left[M^{3}F(R)-\Lambda \right]+\int d^{4}x \sqrt{-g_{i}}V_{i},
\end{equation}
where $\Lambda$ is the bulk cosmological constant, $R$ is the five-dimensional Ricci scalar
and $V_{i}$ is the brane tension for the $i$th brane. A general warped metric ansatz is taken as
\begin{equation}\label{branef13}
ds^{2}=e^{-2A(y)}g_{\mu \nu}dx^{\mu}dx^{\nu}+r_{c}^{2}dy^{2}.
\end{equation}
The Einstein equations with constant scalar curvature for this metric are without
any higher derivative term and therefore free from ghosts. They are
\begin{eqnarray}\label{branef15}
\left\lbrace ^{4}R_{\mu \nu}+\frac{e^{-2A}}{r_{c}^{2}}
\left(A''-4A'^{2}\right)g_{\mu \nu}\right\rbrace \frac{dF(R)}{dR}
\nonumber
\\
-\frac{1}{2}g_{\mu \nu} e^{-2A}F(R)= -\frac{\Lambda}{2M^{3}}e^{-2A}g_{\mu \nu}
\end{eqnarray}
\begin{equation}\label{branef16}
\frac{4}{r_{c}^{2}}\left(A''-A'^{2}\right)\frac{dF(R)}{dR}-\frac{1}{2}F(R)= -\frac{\Lambda}{2M^{3}},
\end{equation}
where the prime denotes derivative with respect to $y$. The five-dimensional scalar curvature
has the expression $R=e^{2A}(^{4}R) +\frac{1}{r_{c}^{2}}\left(8A''-20A'^{2}\right)$.
These equations have to be supplemented by the boundary conditions
$\left[A'(y)\right]_{i}=\frac{\epsilon _{i}}{12M^{3}}V_{i}$, where $\epsilon _{hid}=-\epsilon _{vis}=1$.

From Eqs. (\ref{branef15}) and (\ref{branef16}) we can 
separate the spacetime part, and the extra dimension part which finally leads to
\begin{eqnarray}\label{branef20}
^{4}R_{\mu \nu}&=&\Omega g_{\mu \nu}
\nonumber
\\
3A''&=&\Omega r_{c}^{2}e^{2A}.
\end{eqnarray}
Note that the first equation can equivalently be written as $^{4}G_{\mu \nu}=-\Omega g_{\mu \nu}$. 
From Eq. (\ref{branef15}) we readily obtain a simplified version of equation satisfied by
$A'$ by introducing $F(R)=R+f(R)$, which finally leads to (with rescaling $y\rightarrow r_{c}y$),
\begin{eqnarray}\label{branef22a}
(A')^{2}\Big(6&-&4\frac{df}{dR}\Big)
\nonumber
\\
&=&-\frac{\Lambda}{2M^{3}}+\frac{f}{2}+2\Omega e^{2A} \left(1-\frac{2}{3}\frac{df}{dR}\right).
\end{eqnarray}
The above equation has the following solution for the variable $A$,
\begin{eqnarray}
e^{-A}&=&\omega \cosh \left(\ln \frac{\omega}{c_{1}}+k_{F}y\right)
\label{branef22b}
\\
k_{F}^{2}&=&-\frac{1}{6}\left(\frac{\frac{\Lambda}{2M^{3}}
-\frac{f}{2}}{1-\frac{2}{3}\frac{df}{dR}}\right),
\label{branef22c}
\end{eqnarray}
where $\omega ^{2} \equiv-\frac{\Omega}{3k^{2}}>0$, since for AdS bulk $\Lambda <0$
the induced cosmological constant $\Omega$ on the visible brane is negative. Also note
that since the five-dimensional scalar curvature is constant, the quantity $k_{F}^{2}$ is
actually constant. The respective brane tensions are being given by
\begin{eqnarray}\label{branef22ca}
V_{vis}&=&12M^{3}k_{F}\left[\frac{\frac{\omega ^{2}}{c_{1}^{2}}e^{2kr_{c}\pi}-1}
{\frac{\omega ^{2}}{c_{1}^{2}}e^{2kr_{c}\pi}+1}\right]
\nonumber
\\
V_{hid}&=&12M^{3}k_{F}
\left[\frac{1-\frac{\omega ^{2}}{c_{1}^{2}}}{1+\frac{\omega ^{2}}{c_{1}^{2}}}\right].
\end{eqnarray}
The quantity $c_{1}$ is obtained by normalizing the warp factor to unity
at $y=0$ and leads to $c_{1}=1+\sqrt{1-\omega ^{2}}$. The solution to the 
hierarchy problem can be obtained by calculating the warp factor at $y=\pi r_{c}$.
Following the procedure used in \cite{Sengupta} we also observe that in this scenario we
have both positive and negative cosmological constants on the visible 3-brane.
Remarkably for the anti-de Sitter brane the hierarchy problem can be solved with both the branes having 
positive tension. In all these cases if $k_{F}^{2}>k_{RS}^{2}$, then the brane
tensions are greater than that in the RS model [see Eq. (\ref{branef22ca})]. 
Moreover, if we want to resolve the fine-tuning problem without
introducing any further hierarchy the brane cosmological constant $\omega$ must be very small.

Thus our result is identical to that of the RS model except for the fact that it differs in one
important aspect, the quantity $k_{F}^{2}$ is different and depends on the $f(R)$
model we are considering. On the visible brane as in the RS case, the Higgs mass gets
warped by the following form,
\begin{equation}\label{branef10}
m=m_{0}e^{-2k_{F}r_{c}\pi}
\end{equation}
Since $k_{F}$ is different from the RS solution $r_{c}$ should be modified
such that the product $k_{F}r_{c}$ remains $\approx$ 12 to produce desired warping.

The graviton KK spectrum in these brane-world models acts as an important 
phenomenological signature of extra dimensions.
This is quiet different from the usual factorizable geometry and results in a distinctive phenomenology.
All these KK modes have masses and couplings in the TeV scale, which should be produced on resonance and
hence observed in a TeV-scale collider. As in the RS scenario the metric fluctuation 
can be taken as a linear expansion of the flat
metric around the Minkowski value,
$\widehat{G}_{\alpha \beta}=e^{-2\sigma}\left(\eta _{\alpha \beta}+\kappa ^{*}h_{\alpha \beta} \right)$. 

Though we have been working in constant curvature slices, due to introduction 
of the perturbation $h_{\alpha \beta}$, this will not be true in general. 
Hence we need to consider higher derivative terms which manifest as a scalar degree 
of freedom. However we shall work with a gauge choice, very often used in literature 
to simplify the linearized Einstein's equation \cite{Davoudiasl,Padmanabhan}, such that 
$\partial _{\alpha}h^{\alpha \beta}=0=h^{\alpha}_{\alpha}$. Then the Ricci tensor has the 
following expression:
\begin{eqnarray}\label{new01}
R_{\alpha \beta}&=&-\frac{1}{2}\square h_{\alpha \beta}+\frac{1}{2}\Big\lbrace -\frac{1}{r_{c}^{2}}
\partial _{y}\left(e^{-2\sigma}\partial _{y}h_{\alpha \beta}\right)
\nonumber
\\
&+&\frac{2}{r_{c}^{2}}e^{-2\sigma}\frac{d\sigma}{dy}\partial _{y}h_{\alpha \beta}\Big\rbrace
-\frac{4}{r_{c}^{2}}e^{-2\sigma}\left(\frac{d\sigma}{dy}\right)^{2}\eta _{\alpha \beta}
\nonumber
\\
&+&\frac{1}{r_{c}^{2}}\frac{d^{2}\sigma}{dy^{2}}e^{-2\sigma}
\left(\eta _{\alpha \beta}+h_{\alpha \beta}\right)
\\
R_{yy}&=&4\frac{d^{2}\sigma}{dy^{2}}-4\left(\frac{d\sigma}{dy}\right)^{2}.
\end{eqnarray}
Thus on imposing those gauge conditions and the results 
$\left(d\sigma /dy\right)^{2}=\left(k_{F}r_{c}\right)^{2}$ and 
$d^{2}\sigma /dy^{2}=2k_{F}r_{c}\left[\delta (y)-\delta (y-\pi)\right]$, the scalar curvature 
turns out to be $R=-16k_{F}^{2}$, a constant. Therefore, under these gauge conditions the 
bulk curvature is constant and we can put the above results directly 
into the higher derivative free Einstein's equation (\ref{branef15}).
The metric perturbation leading to graviton KK modes can be expressed as\cite{Davoudiasl}
\begin{equation}\label{branegrav1}
h_{\alpha \beta}(x_{\alpha},y)=\sum _{n=0}^{\infty}h_{\alpha \beta}^{(n)}(x_{\alpha})
\frac{\chi ^{(n)}(y)}{\sqrt{r_{c}}},
\end{equation}
where $h^{(n)}_{\alpha \beta} (x_{\alpha})$ corresponds to KK modes of the four-dimensional graviton.
The equation satisfied by the mode functions $\chi ^{(n)}$ is given by
\begin{equation}\label{branegrav2}
\frac{-1}{r_{c}^{2}}\frac{d}{dy}\left(e^{-4\sigma} \frac{d\chi ^{(n)}}{dy}\right)
=m_{n}^{2}e^{-2\sigma}\chi ^{n},
\end{equation}
along which the equation of motion for the graviton modes 
$h^{(n)}_{\alpha \beta}(x_{\mu})$ has 
the following expression:
\begin{equation}\label{branenew02}
\left(n^{\mu \nu}\partial _{\mu}\partial _{\nu}-m_{n}^{2}\right)h^{(n)}_{\alpha \beta}(x_{\gamma})=0.
\end{equation}
Note that Eq. (\ref{branegrav2}) is actually modified by the introduction of 
the $F(R)$ term through the factor $\sigma$ which contains the factor $k_{F}$, which depends 
explicitly on the modified gravity terms [see Eq. (\ref{branef22c})].
The functions $\chi ^{(n)}$ satisfy the orthonormality condition, 
$\int _{-\pi}^{\pi}d\phi e^{-2\sigma}\chi ^{(m)}\chi ^{(n)}=\delta _{nm}$.
The solutions for $\chi ^{(n)}$ are being given by \cite{Goldberger2},
\begin{equation}\label{branegrav3}
\chi ^{(n)}(y)=\frac{e^{2\sigma (y)}}{N_{n}}\left[J_{2}(z_{n})+\alpha _{n}Y_{2}(z_{n})\right]
\end{equation}
where $J_{2}$ and $Y_{2}$ are Bessel functions of second order, $z_{n}(y)=m_{n}e^{\sigma (y)}/k$
and $N_{n}$ is the normalization. Then for $x_{n}=z_{n}(\pi)$ and in the limit
$m_{n}/k\ll 1$ and $e^{kr_{c}\pi}\gg 1$, using continuity of the first derivative of $\chi ^{(n)}$
we get $x_{n}$ to be the solution of the equation $J_{1}(x_{n})=0$. The masses of the graviton
KK excitations are given by $m_{n}=k_{F}x_{n}e^{-k_{F}r_{c}\pi}$.

Having obtained the mass modes, we can now derive the interactions of
$h_{\alpha \beta}$ with the matter fields on the visible 3-brane at $\phi =\pi$.
We find the usual form of the interaction Lagrangian as
\begin{equation}\label{branegrav3a}
L=−\frac{1}{M^{3/2}}T^{\alpha \beta}(x)h_{\alpha \beta}(x,\phi =\pi),
\end{equation}
where $T_{\alpha \beta}(x)$ is the symmetric energy-momentum tensor of the matter fields.
Expanding the graviton field into the KK states of Eq. (\ref{branegrav1}) and using
the proper normalization for $\chi _{n}(\phi)$ we arrive at
\begin{equation}\label{branegrav3b}
L=−\frac{1}{M_{Pl}}T^{\alpha \beta}h_{\alpha \beta}^{(0)}
-\frac{1}{\Lambda _{\pi}}\sum _{n=1}^{\infty}T^{\alpha \beta}h_{\alpha \beta}^{(n)}.
\end{equation}
Thus, as in the RS scenario, the zero mode in $F(R)$ gravity 
couples with the inverse four-dimensional Planck scale 
where all the massive KK states couple as $\Lambda _{\pi}^{-1}$, 
where $\Lambda _{\pi} = e^{−k_{F}r_{c}\pi} M_{Pl}$, 
which is of order $\textrm{TeV}^{-1}$
i.e. in the weak scale. Thus even in the $F(R)$ 
theory the couplings are the same as in the RS model.
Hence the 3 TeV experimental lower mass bound on 
first graviton KK modes applies here also.

The null result in LHC results in a lower mass bound for KK modes of gravitation which is 3 TeV. 
However as discussed earlier, it is difficult to achieve 
this in the standard RS model. Following our calculation due
to inclusion of $F(R)$ gravity in the bulk, we readily obtain that the graviton masses
become modified by the factor $k_{F}$ given by Eq. (\ref{branef22c}). 
However in an $F(R)$ model if $k_{F}/k_{RS}=p$
then retaining $k_{F}r_{c}\sim 12$, one can set the graviton mass $p$ times
larger than the RS model to raise it above the experimental lower bound.

Thus if we take $k_{F}^{2}=-p\frac{\Lambda}{12M^{3}}$ where $p>1$ we arrive at the following relation,
\begin{equation}\label{branegrav4}
 p\frac{df}{dR}\frac{\Lambda}{18M^{3}}=(p-1) \frac{\Lambda}{12M^{3}}+\frac{f}{12}.
\end{equation}
For a particular choice $f(R)=\beta R^{n}$, we find
\begin{equation}\label{branegrav5}
\frac{\beta R^{n-1}}{12}\left(\frac{2pn\mid \Lambda \mid}{3M^{3}}+R\right)
=(p-1)\frac{\mid \Lambda \mid}{12M^{3}}.
\end{equation}
Thus we get a bound on the scalar curvature given by
\begin{eqnarray}\label{branegrav6}
R&>&-\frac{2pn\mid \Lambda \mid}{3M^{3}}
\nonumber
\\
(\beta &>&0, \textrm{n=odd};~ \beta <0, \textrm{n=even})
\nonumber
\\
R&<&-\frac{2pn\mid \Lambda \mid}{3M^{3}}
\nonumber
\\
(\beta &>&0, \textrm{n=even};~ \beta <0, \textrm{n=odd}).
\end{eqnarray}
Now we would like to know what are the criteria for $k_{F}^{2}$ to be greater than $k_{RS}^{2}$.
In the limit $f(R)\rightarrow 0$ we readily obtain
$k_{F}^{2}=k_{RS}^{2}=\frac{-\Lambda}{12M^{3}}$. Then requiring 
$k_{F}^{2} > \frac{\mid \Lambda \mid}{12M^{3}}$
we arrive at the following criteria:
\begin{equation}\label{branef9b}
\frac{f(R)}{df/dR}>-\frac{2\mid \Lambda \mid}{3M^{3}}.
\end{equation}
From Eq. (\ref{branef9b}), with $f(R)=\beta R^{n}$ we obtain the bound
\begin{equation}\label{branegrav7}
\mid R\mid <\frac{2n\mid \Lambda \mid}{3M^{3}}.
\end{equation}
If we employ the criteria that $k_{F}^{2}>0$ then we shall also retrieve the criteria given by
Eq. (\ref{branegrav7}). However the criteria $k_{F}/M<1$ 
(i.e. bulk curvature is less than five-dimensional Planck scale
to ensure that the classical solutions of Einstein's equation 
can be trusted \cite{LHC},\cite{Davoudiasl}) leads to an inequality given by
\begin{equation}\label{branegrav8}
\frac{\mid \Lambda \mid}{12M^{5}}+\frac{\beta R^{n-1}}{2M^{2}}
\left\lbrace R+\frac{4}{3}nM^{2}\right\rbrace <1.
\end{equation}
As the bulk is AdS, we infer that the five-dimensional scalar curvature 
must be negative. This implies $\beta >0$
is the valid choice for odd powers of R, while $\beta <0$ is a valid choice for
even powers of R. Hence the form of $F(R)$ should be,
$F(R)=R-\alpha R^{2}+\beta R^{3}-\cdots$. From these we 
arrive at the following bound on the bulk curvature:
\begin{equation}\label{branegrav9}
|R|<\textrm{min}\left(\frac{2n\mid \Lambda \mid}{3M^{3}},\left[\frac{3}{2n\beta}
\left(1-\frac{\mid \Lambda \mid}{12M^{5}}\right)\right]^{1/n-1}\right).
\end{equation}
The above inequality exhibits that the magnitude of the bulk curvature
should remain less than $\mathcal{O}\left(\mid \Lambda \mid/12M^{3}\right)$. 

We can go further with the help of Ref. \cite{Pogosian} to discuss bounds 
on the parameters $\alpha$, $\beta$ in the proposed $F(R)$ gravity model. These 
stringent conditions appear in order to make the model consistent with experimental 
and observed data relating to cosmic expansion. Thus we discuss all the conditions as 
presented in Ref. \cite{Pogosian} to see for what ranges of $\alpha$ and $\beta$ the 
model becomes viable.\\
$~~~~$First, the condition $1+\frac{df}{dR}>0$ 
for all finite $R$ implies that the effective Newton's 
constant does not change sign, while from a microscopic level this implies the graviton 
does not turn to ghost modes \cite{Nariai,Nunez}. This leads to the condition 
$2\alpha -3\beta \mid R \mid < \mid R \mid ^{-1}$. Thus if $\beta =0$ then we 
can constrain the parameter $\alpha$ to be
$\alpha < \left(2\mid R \mid \right)^{-1} \sim 6M^{3}\mid \Lambda \mid ^{-1}$.\\
$~~~~~$Next, we will impose the condition $df/dR<0$. This comes from tight 
constraints related to big bang nucleosynthesis and the cosmic microwave background. 
This condition amounts to setting, $2\alpha >3\beta \mid R \mid$. Thus for the constraint 
on $\alpha$ from the previous condition, we obtain the following bound on the 
parameter $\beta$: $\beta < 0.33\left(12M^{3}\right)^{2}\mid \Lambda \mid ^{-2}$.\\
$~~~~~$Finally, we consider the situation where $d^{2}f/dR^{2}>0$ for higher values 
of $R$. This ensures that scalaron modes are not tachyonic. This condition along with those 
mentioned earlier, leads to a bound of the form $6\beta \mid R \mid >2\alpha >3\beta \mid R \mid$. 
Also we should mention that recent galaxy formation surveys have 
constrained $df/dR$ to be smaller than $10^{-6}$ \cite{Hu}. However, such 
statements are yet to be confirmed  since galaxy formation has not been studied 
in the $F(R)$ model using N-body simulations.\\
$~~~~$We have thus shown that although the graviton KK mode mass bound in recent
LHC experiments cannot be explained in the standard RS scenario, it can be explained
quiet well in the $F(R)$ gravity framework. The solutions on the branes for the nonflat
metric ansatz have been obtained with $F(R)$ gravity in the bulk, which shows that
though they have the features similar to the RS scenario, they differ in one important aspect.
The parameter, $k_{F}$, depends on the form of $F(R)$. From the flat
brane limit we recover the RS-like solutions in $F(R)$ gravity on the bulk.
The graviton KK modes get modified by the factor $k_{F}$ and can have
values well beyond $3$ TeV, without any conflict with recent LHC results.
Finally, from various criteria like $k_{F}^{2}>0$, $k_{F}/M<1$, we arrive at
a constraint equation on the bulk curvature. By assuming some reasonable
properties about the bulk curvature (for example, it should be $\mathcal{O} |\Lambda| /12M^{3})$ we
observe from Eq. (\ref{branegrav9}) that this holds, provided $F(R)$ has a leading order term as
$n=2$ with a negative coefficient. Also, by considering viable 
forms for the $f(R)$ model, we have constrained the parameters of our model in terms 
of basic variables in our theory.\\
$~~~~$Hence, by introducing $F(R)$ gravity in the bulk, we have obtained a 
positive-tension visible brane with small cosmological constant. Moreover,
we have also derived graviton mass modes and couplings with the possibility of 
having the mass of first graviton KK modes above 3 TeV, 
so that the model survives recent ATLAS results 
at the LHC. Finally some constraints on bulk scalar 
curvature are obtained in these higher curvature gravity 
models.

\section*{Acknowledgements}

S.C. is funded by a SPM fellowship from CSIR, Government of India.


\end{document}